\documentclass[amsmath,amssymb,twocolumn]{revtex4}
\usepackage{graphicx}
\usepackage{amscd,amsmath,amsthm,amsfonts,amssymb}

\def\sech{{\rm sech}}
\def\tanh{{\rm tanh}}

\def\ri{{\rm i}}

\def\re{{\rm e}}
\def\PT{$\mathcal{PT}$}
\def\P{\mathcal{P}}

\begin{document}
\title{New classes of non-parity-time-symmetric optical potentials with all-real spectra and exceptional-point-free phase transition}
\author{Jianke Yang}
\address{Department of Mathematics and Statistics, University of Vermont, Burlington, VT 05401, USA}

\begin{abstract}
Paraxial linear propagation of light in an optical waveguide with material gain and loss is governed by a Schr\"odinger equation with a complex potential. Properties of parity-time-symmetric complex potentials have been heavily studied before. In this article, new classes of non-parity-time-symmetric complex potentials featuring conjugate-pair eigenvalue symmetry in its spectrum are constructed by operator symmetry methods. Due to this eigenvalue symmetry, it is shown that the spectrum of these complex potentials is often all-real. Under parameter tuning in these potentials, phase transition can also occur, where pairs of complex eigenvalues appear in the spectrum. A peculiar feature of the phase transition here is that, the complex eigenvalues may bifurcate out from an interior continuous eigenvalue inside the continuous spectrum, in which case a phase transition takes place without going through an exceptional point. In one spatial dimension, this class of non-parity-time-symmetric complex potentials is of the form $V(x)=h'(x)-h^2(x)$, where $h(x)$ is an arbitrary parity-time-symmetric complex function. These potentials in two spatial dimensions are also derived. Diffraction patterns in these complex potentials are further examined, and unidirectional propagation behaviors are demonstrated.
\end{abstract}

\maketitle

Parity-time (\PT) symmetric optics has been heavily studied in the past ten years (see \cite{Kivshar_review,Yang_review} for reviews). \PT symmetry was first introduced as a non-Hermitian generalization of quantum mechanics, where a complex but \PT-symmetric potential was shown to still possess all-real spectrum \cite{Bender1998}. Since the Schr\"odinger equation in quantum mechanics is equivalent to the beam propagation equation in optics under paraxial approximation, \PT symmetry then spread to optics, where this symmetry can be realized by an even refractive index profile together with an odd gain-loss landscape \cite{Musslimani2008}. In this optical setting, \PT symmetry was studied experimentally for the first time \cite{Ruter_2010,Regensburger_2012}. \PT-symmetric potentials do not always possess all-real spectra though. If the gain-loss profile (corresponding to the imaginary part of the complex potential) is too strong, phase transition can occur, where complex eigenvalues enter the spectrum \cite{Bender1998,Ahmed2001}. This phase transition has been demonstrated in a number of optical experiments and utilized for many emerging applications \cite{Ruter_2010,Kottos,Regensburger_2012,Feng2013,Peng,PTlaser_Zhang,PTlaser_CREOL}. The interplay between \PT symmetry and nonlinearity has been extensively explored as well \cite{Kivshar_review,Yang_review}.

Optical \PT symmetry requires the refractive index to be even and gain-loss profile to be odd in space, which is restrictive and limiting its applicability. Generalization of \PT-symmetric potentials to allow a more flexible gain-loss profile while still maintaining all-real spectrum is thus an important question. Using supersymmetry methods, families of non-\PT-symmetric potentials with real spectra have been constructed \cite{Cannata1998,SUSY2013}. But the gain-loss profiles in such potentials are still very special. Using two other methods, classes of non-\PT-symmetric potentials with arbitrary gain-loss profiles and all-real spectra were reported in \cite{Tsoy2014,Yang2016}, which is a big step forward. The simplest class of such potentials is of the form $V(x)=g^2(x)+ig'(x)$, where $g(x)$ is an arbitrary real function. Other such potentials with more involved functional forms can be found in \cite{Yang2016}.

In this article, we report new classes of non-\PT-symmetric complex potentials which feature all-real spectra and phase transition. In one dimension, this class of potentials is of the form $V(x)=h'(x)-h^2(x)$, where $h(x)$ is an arbitrary \PT-symmetric complex function. Even though these potentials are very different from those reported in \cite{Tsoy2014,Yang2016}, we show that their eigenvalues still possess conjugate-pair symmetry. Thus, their spectrum can still be all-real, and phase transition can also occur by tuning parameters in these potentials. Since $h(x)$ is an arbitrary \PT-symmetric function, the gain-loss profile in these potentials can also be arbitrary.
A peculiar phenomenon here is that, when a phase transition occurs in a localized potential, discrete complex eigenvalues can bifurcate out from a continuous real eigenvalue in the interior of the continuous spectrum. When this happens, the phase transition does not go through an exceptional point. This scenario of a phase transition is different from all those reported before in non-Hermitian systems, where a phase transition was induced by collisions of real eigenvalues through an exceptional point \cite{Bender1998,Ahmed2001,Ruter_2010,Kottos,Regensburger_2012,Peng,PTlaser_Zhang,PTlaser_CREOL,Yang_review,Yang2016}. For periodic potentials at phase transition, we observe unidirectional beam propagation behaviors with new features beyond those reported earlier for \PT-symmetric potentials.

Linear paraxial propagation of light in an optical waveguide with material gain and loss is governed by the Schr\"{o}dinger equation
\begin{equation} \label{e:SE}
\ri \Psi_z + \Psi_{xx} + V(x) \Psi=0,
\end{equation}
where $z$ is the distance of propagation, $x$ is the transverse coordinate,
$V(x)$ is a complex potential whose real part is the index of refraction and whose imaginary part represents gain and loss in the waveguide. Looking for eigenmodes of the form $\Psi = \re^{-\ri \mu z} \psi(x)$ we arrive at the eigenvalue problem
\begin{equation} \label{e:Lpsi}
L\psi=-\mu \psi,
\end{equation}
where
$L = \partial_{xx} + V(x)$ is a Schr\"odinger operator and $\mu$ is an eigenvalue.

To derive new non-\PT-symmetric complex potentials with all-real spectra, we follow the operator symmetry strategy which we have used before \cite{Yang2016}. This strategy is based on the following observation: if there exists an operator $\eta$ such that $L$ and its complex conjugate $L^*$ are related by a similarity relation
\begin{equation} \label{Eq:LSimilar}
\eta L =  L^* \eta,
\end{equation}
then the eigenvalues of $L$ come in conjugate pairs if the kernel of $\eta$ is empty. This conjugate-pair eigenvalue symmetry guarantees that either the spectrum of $L$ is all-real, or a phase transition occurs when pairs of complex eigenvalues bifurcate out.

The key difference between our current approach and the one in \cite{Yang2016} is that, instead of choosing $\eta$ as pure differential operators, we now take $\eta$ to be a combination of the parity operator $\P$ and differential operators. In one spatial dimension, the parity operator is defined as $\P f(x)\equiv f(-x)$. In the simplest case, we take $\eta$ to be a combination of the parity operator and a first-order differential operator, i.e.,
\begin{equation} \label{e:eta1}
\eta =\P \left[\partial_x + h(x)\right],
\end{equation}
where $h(x)$ is a complex function to be determined. Substituting this $\eta$ into the
similarity condition (\ref{Eq:LSimilar}), we get the following two
equations
\begin{equation} \label{e:vpv}
V(x)-V^*(-x)=2h'(x),
\end{equation}
\begin{equation} \label{e:VV}
\left[V(x)-V^*(-x)\right]h(x)=h''(x)-V'(x).
\end{equation}
From the first equation, we see that $\left[h^*(-x)\right]_x=h'(x)$; thus
\begin{equation} \label{e:h}
h^*(-x)=h(x)+c_1,
\end{equation}
where $c_1$ is a constant. Substituting Eq. (\ref{e:vpv}) into (\ref{e:VV}) and integrating once, we get
\begin{equation} \label{e:Vx}
V(x)=h'(x)-h^2(x)+c_2,
\end{equation}
where $c_2$ is another constant. Lastly, inserting (\ref{e:h}) and (\ref{e:Vx}) into Eq. (\ref{e:vpv}), we obtain
\begin{equation} \label{e:c1c2}
c_1^2+2c_1h(x)+c_2-c_2^*=0.
\end{equation}
In order for the potential $V(x)$ in (\ref{e:Vx}) not to be a constant, the function $h(x)$ should not be identically zero. Thus, Eq. (\ref{e:c1c2}) dictates that $c_1=0$ and $c_2$ is real. The former condition means that the complex function $h(x)$ is \PT-symmetric in view of Eq. (\ref{e:h}). Regarding the latter condition, since a real constant in a potential can be easily removed by a simple shift of the eigenvalue, we can set $c_2=0$ without loss of generality. In the end, we find that for new complex potentials of the form
\begin{equation} \label{e:V}
V(x)=h'(x)-h^2(x),
\end{equation}
where $h(x)$ is a \PT-symmetric complex function, i.e., $h^*(x)=h(-x)$, the Schr\"odinger operator $L$ satisfies the similarity condition (\ref{Eq:LSimilar}) with $\eta$ given in (\ref{e:eta1}). Because of this, eigenvalues of $L$ exhibit complex-conjugate symmetry; hence
the spectrum of $L$ can be all-real and phase transition could occur, similar to \PT-symmetric potentials.

In the above new potentials, $h(x)$ is an arbitrary \PT-symmetric function. This implies that the imaginary part of the potential $V(x)$ (corresponding to the gain and loss profile) can also be arbitrary. To see this, we write $h(x)=h_1(x)+ih_2(x)$, where $h_1$ is the real part of $h$ which is even and $h_2$ the imaginary part of $h$ which is odd. Then, $\mbox{Im}(V)=h'_2(x)-2h_1(x)h_2(x)$. Notice that $h_2'(x)$ is even and $h_1(x)h_2(x)$ is odd. For any arbitrary gain-loss profile $G(x)=\mbox{Im}(V)$, we can always write it as $G(x)=G_1(x)+G_2(x)$, where $G_1(x)\equiv [G(x)+G(-x)]/2$ is even and $G_2(x)\equiv [G(x)-G(-x)]/2$ is odd. Then under the choice of $h'_2(x)=G_1(x)$ and $h_1(x)=-G_2(x)/2h_2(x)$, where $h_2(x)=\int G_1(x)dx$ is selected to be an odd function, the resulting gain-loss profile $\mbox{Im}(V)=h'_2(x)-2h_1(x)h_2(x)$ would be equal to $G(x)$. This means that the new class of potentials (\ref{e:V}) can accommodate any arbitrary gain-loss profile (the refractive index would need to be engineered accordingly though).

In the above construction, if we choose $\eta$ to be a combination of the parity operator $\P$ and higher-order differential operators, additional families of new non-\PT-symmetric complex potentials with all-real spectra could be derived.

To illustrate the all-real spectra and phase transition of this class of non-\PT-symmetric potentials, we take
\begin{equation} \label{e:hb}
h(x)=b_0+b_1\cos x+ib_2\sin x,
\end{equation}
which is a periodic function with real constants $b_0, b_1$ and $b_2$. This function $h(x)$ is \PT-symmetric as required. We also fix $b_0=b_1=1$ and allow $b_2$ to vary. When $b_2=0.98$, the resulting complex potential $V(x)$ is depicted in the upper left panel of Fig. 1. It is easy to see that this potential is non-\PT-symmetric (even under any $x$-coordinate shift). The diffraction relation of this periodic potential is shown in the upper right panel of Fig. 1. These diffraction curves are all-real, meaning that the whole spectrum of $L$ is all-real. However, when $b_2=1.02$, the lowest two Bloch bands collide and complex eigenvalues appear near the two edges of the Brillouin zone, see the lower panels of Fig. 1. In this case, the spectrum of $L$ becomes partially complex. The phase transition occurs at $b_2=1$.
These behaviors are qualitatively similar to that in \PT-symmetric periodic potentials \cite{Musslimani2008}.

\begin{figure}[tb!]
\includegraphics[width=0.48\textwidth]{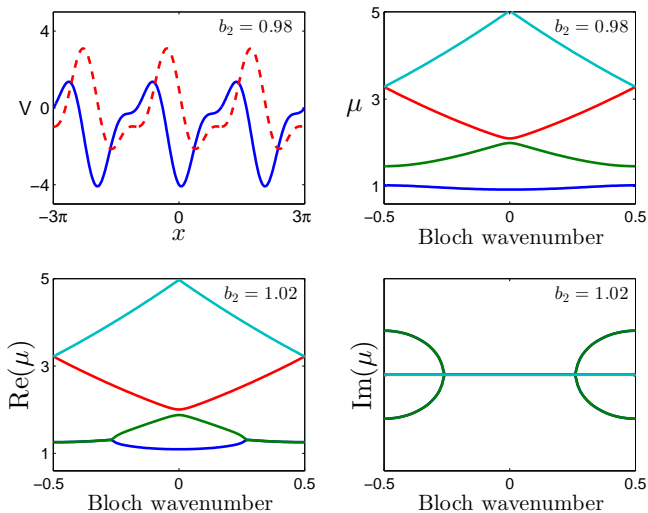}

\smallskip
\caption{Diffraction relations of non-\PT-symmetric periodic potentials (\ref{e:V}) with $h(x)$ given in (\ref{e:hb}) and $b_0=b_1=1$. Upper left: real (solid blue) and imaginary (dashed red) parts of the complex potential at $b_2=0.98$. Upper right: diffraction curves of the potential in the upper left panel over the Brillouin zone $-0.5\le k\le 0.5$. Lower panels: real and imaginary parts of the diffraction curves over the Brillouin zone at $b_2=1.02$. }
\label{Fig1}
\end{figure}

Dynamics of beam propagation in these new complex potentials is of high interest. To examine this, we take the above periodic potential at phase transition ($b_2=1$), and launch a broad beam into it at opposite angles. Specifically, our initial condition is taken as
\begin{equation} \label{e:ic}
\Psi(x,0)=e^{-x^2/100+ i\beta x},
\end{equation}
where the real constant $\beta$ is proportional to the initial launch angle. When $\beta=\pm 2$ (opposite angles), evolutions of these two beams are obtained by computing Eq. (\ref{e:SE}) and displayed in Fig. 2. We find that when the beam is launched toward the left, it does not really travel in that direction. Instead, it spreads in both directions (see the left panel). On the other hand, if the beam is launched to the right, it indeed moves away toward that direction. Thus, this non-\PT-symmetric periodic potential exhibits highly-nonreciprocal unidirectional behavior. While non-reciprocity and unidirectional propagation have been reported in \PT-symmetric photonic lattices before \cite{Longhi,Musslimani2010,Ruter_2010,Regensburger_2012,Feng2013}, this behavior in the underlying non-\PT-symmetric lattices is worthy of report. In addition, details of these nonreciprocal unidirectional behaviors here are not exactly the same as those reported earlier for \PT-symmetric lattices.

\begin{figure}[tb!]
\includegraphics[width=0.48\textwidth]{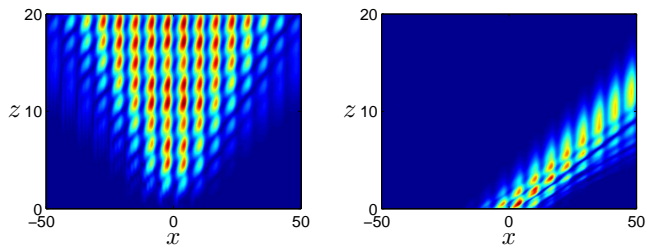}

\smallskip
\caption{Propagation of broad beams (\ref{e:ic}) launched at opposite angles into the non-\PT-symmetric lattice (\ref{e:V}) and (\ref{e:hb}) at phase transition ($b_2=1$). Left: $\beta=-2$; right: $\beta=2$. }
\label{Fig2}
\end{figure}

The previous example was a periodic potential induced by a periodic function $h(x)$. When $h(x)$ is chosen as a localized function, a localized non-\PT-symmetric potential would result. As an example, we take
\begin{equation} \label{e:hlocal}
h(x)=d_1 \hspace{0.02cm} \sech x+i \hspace{0.02cm} d_2 \hspace{0.03cm} \sech x \hspace{0.04cm} \tanh x,
\end{equation}
which is \PT-symmetric for real constants $d_1$ and $d_2$. We also fix $d_1=1$. Then for two different $d_2$ values of 1 and 2, the resulting non-\PT-symmetric potentials and their linear spectra are plotted in Fig. 3. Since these potentials are localized, their continuous spectra are the same, which are $0\le \mu< \infty$. When $d_2=1$, there are no discrete eigenvalues; thus the spectrum is all-real (see the upper right panel). But when $d_2=2$, a conjugate pair of discrete eigenvalues $\mu\approx 0.7067\pm 0.4961i$ appear (see the lower right panel). The phase transition occurs at $d_2\approx 1.385$. The most fascinating feature of this phase transition is that, the two complex eigenvalues bifurcate out from $\mu_0\approx 0.8062$, which is in the interior of the continuous spectrum. We also noticed that the discrete (localized) eigenfunctions of the two complex eigenvalues bifurcate out from two different continuous (nonlocal) eigenfunctions of the real eigenvalue $\mu_0$. This reveals two facts: (1) these discrete eigenmodes bifurcate out from continuous eigenmodes, rather than embedded isolated eigenmodes, in the interior of the continuous spectrum; (2) this phase transition does not go through an exceptional point. The second fact is particularly significant because, to our knowledge, all phase transitions reported before in both finite- and infinite-dimensional non-Hermitian systems occurred due to a collision of real eigenvalues forming an exceptional point, where different eigenvectors or eigenfunctions coalesce   \cite{Bender1998,Ahmed2001,Yang_review,Yang2016}. This is the first instance where a phase transition occurs without an exceptional point. We also point out that, from a mathematical point of view, we have not seen discrete eigenvalues bifurcating out of a continuous eigenvalue in the interior of the continuous spectrum before, and this is the first such example. Why this bifurcation could happen here is a mathematical mystery.

\begin{figure}[tb!]
\includegraphics[width=0.48\textwidth]{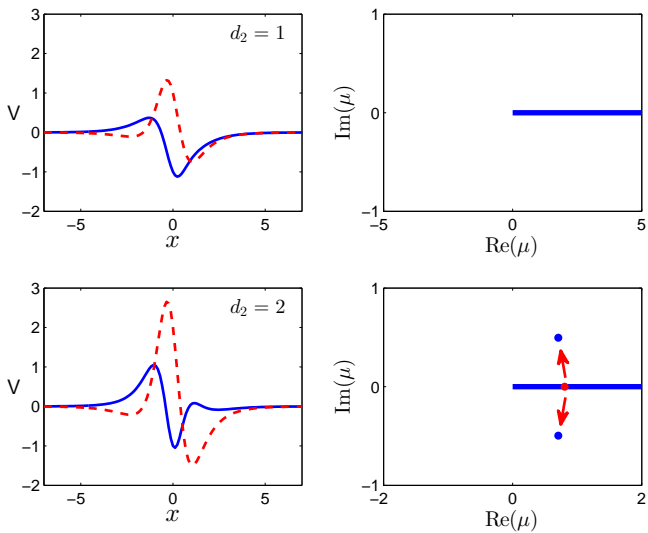}

\smallskip
\caption{Spectra of localized potentials (\ref{e:V}) with $h(x)$ given in (\ref{e:hlocal}) and $d_1=1$ (the $d_2$ values are shown inside the panels). Left column: real (solid blue) and imaginary (dashed red) parts of the complex potentials. Right column: spectra of potentials in the left column (the red arrows in the lower panel indicate that the two complex eigenvalues in the spectrum bifurcate out from the red dot in the interior of the continuous spectrum when a phase transition happens). }
\label{Fig3}
\end{figure}

Lastly, we show that our construction of non-\PT-symmetric complex potentials above can be extended to higher dimensions. Let us
consider paraxial light propagation in a three-dimensional (3D)
waveguide, which gives rise to a 2D Schr\"{o}dinger operator
\begin{equation} \label{e:V2D}
L=\partial_{xx}+\partial_{yy}+V(x,y).
\end{equation}
To construct new complex potentials $V(x,y)$ with all-real spectra, we still impose the similarity condition (\ref{Eq:LSimilar}), and choose the operator $\eta$ to be a combination of the parity operator and differential operators. In 2D, the parity operator $\P$ can take different forms, either a full parity operator $\P f(x,y)=f(-x,-y)$, or a partial parity operator $\P f(x,y)=f(-x,y)$ or $\P f(x,y)=f(x,-y)$ \cite{YangPPT}. For simplicity, we choose $\eta$ to be a combination of one of those parity operators and a first-order differential operator in $x$, i.e.,
\begin{equation}
\eta=\P \left[\partial_x+a(x)\right].
\end{equation}
Inserting this $\eta$ into the similarity condition (\ref{Eq:LSimilar}) and after some algebra, we obtain the resulting 2D complex potential as
\begin{equation}
V(x,y)=a'(x)-a^2(x)+\phi(y),
\end{equation}
where functions $a(x)$ and $\phi(y)$ satisfy the parity conditions
\begin{equation}
\P a(x) = a^*(x), \quad \P \phi(y) = \phi^*(y).
\end{equation}
This separable 2D complex potential satisfies the similarity condition (\ref{Eq:LSimilar}); thus its complex eigenvalues come in conjugate pairs, which implies that its spectrum can be all-real. Other choices of the operator $\eta$ could lead to additional classes of 2D non-\PT-symmetric complex potentials with all-real spectra.

In summary, we have derived new classes of non-\PT-symmetric optical potentials featuring conjugate-pair eigenvalue symmetry in its spectrum by operator symmetry methods. Due to this eigenvalue symmetry, it is shown that the spectrum of these complex potentials is often all-real. Under parameter tuning in these potentials, phase transition can also occur, where pairs of complex eigenvalues appear in the spectrum. A remarkable finding is that a phase transition in these potentials may not go through an exceptional point, which is novel to our knowledge. Since these new potentials allow an arbitrary gain-loss profile, they may find applications such as non-\PT-symmetric lasers with more flexible laser cavities.

This work was supported in part by the Air Force Office of Scientific Research under award number
FA9550-12-1-0244, and the National Science Foundation under award number DMS-1616122.


\begin{thebibliography}{99}

\bibitem{Yang_review}
V.V. Konotop, J. Yang and D.A. Zezyulin,
``Nonlinear waves in PT-symmetric systems",
Rev. Mod. Phys. 88, 035002 (2016).

\bibitem{Kivshar_review}
S.V. Suchkov, A.A. Sukhorukov, J. Huang, S.V. Dmitriev,
C. Lee and Y.S. Kivshar,
``Nonlinear switching and solitons in \PT-symmetric photonic systems",
Laser Photon. Rev. 10, 177 (2016).

\bibitem{Bender1998} { C.M. Bender and S. Boettcher,}
``Real spectra in non-Hermitian Hamiltonians having PT symmetry",
\emph{Phys. Rev. Lett.} 80,
5243--5246 (1998).

\bibitem{Musslimani2008}
{ Z.H. Musslimani, K.G. Makris, R. El-Ganainy and D.N.
Christodoulides,}
``Optical solitons in PT periodic potentials",
\emph{Phys. Rev. Lett.} 100, 030402 (2008).

\bibitem{Ruter_2010}  C.E. R\"uter, K.G. Makris, R. El-Ganainy, D.N. Christodoulides, M. Segev and D.
Kip,
``Observation of parity–-time symmetry in optics",
Nature Physics 6, 192--195 (2010).

\bibitem{Regensburger_2012}
A. Regensburger, C. Bersch, M.A. Miri, G. Onishchukov, D.N.
Christodoulides and U. Peschel,
``Parity–time synthetic photonic lattices",
Nature \textbf{488}, 167--171 (2012).

\bibitem{Ahmed2001} Z. Ahmed,
``Real and complex discrete eigenvalues in an exactly solvable one-dimensional complex PT-invariant potential",
Phys. Lett. A 282, 343--348 (2001).

\bibitem{Kottos}
J. Schindler, A. Li, M.C. Zheng, F.M. Ellis, and T. Kottos,
``Experimental study of active LRC circuits with \PT symmetries",
Phys. Rev. A 84, 040101(R) (2011).

\bibitem{Feng2013}
L. Feng, Y.L. Xu,  W.S. Fegadolli, M.H. Lu, J.E.B. Oliveira, V.R.
Almeida, Y.F. Chen, and A. Scherer,
``Experimental demonstration of a unidirectional reflectionless parity-time metamaterial at optical frequencies",
Nature Materials, 12, 108--113 (2013).

\bibitem{Peng}
B. Peng, S. \"Ozdemir, F. Lei, F. Monifi, M. Gianfreda, G. Long, S. Fan, F. Nori, C.M.
Bender, and L. Yang,
``Parity-time-symmetric whispering-gallery microcavities",
Nat. Phys. 10, 394 (2014).

\bibitem{PTlaser_Zhang} L. Feng, Z.J. Wong, R. Ma, Y. Wang, and X. Zhang,
``Single-mode laser by parity-time symmetry breaking",
Science 346, 972--975 (2014).

\bibitem{PTlaser_CREOL} H. Hodaei, M.-A. Miri, M. Heinrich, D. N. Christodoulides, and M.
Khajavikhan,
``$\mathcal{PT}$-symmetric micro-ring laser",
Science \textbf{346}, 975--978 (2014).

\bibitem{Cannata1998}
F. Cannata, G. Junker, and J. Trost,
``Schr\"odinger operators with complex potential but real spectrum",
Phys. Lett. A 246, 219--226 (1998).

\bibitem{SUSY2013}
{M.A. Miri, M. Heinrich, and D. N. Christodoulides,}
``Supersymmetry-generated complex optical potentials with real spectra",
\emph{Phys. Rev. A} 87, 043819 (2013).

\bibitem{Tsoy2014} {E.N. Tsoy, I.M. Allayarov and F. Kh.
Abdullaev,}
``Stable localized modes in asymmetric waveguides with gain and loss",
\emph{Opt. Lett.} 39, 4215--4218 (2014).

\bibitem{Yang2016}
S. Nixon and J. Yang,
``All-real spectra in optical systems with arbitrary gain and loss distributions,"
Phys. Rev. A 93, 031802(R) (2016).

\bibitem{Longhi}
S. Longhi,
``Bloch Oscillations in Complex Crystals with \PT Symmetry",
Phys. Rev. Lett. 103, 123601 (2009).

\bibitem{Musslimani2010}
K.G. Makris, R. El Ganainy, D.N. Christodoulides, and Z.H. Musslimani,
``\PT-symmetric optical lattices",
Phys. Rev. A 81, 063807 (2010).

\bibitem{YangPPT}
J. Yang,
``Partially PT-symmetric optical potentials with all-real spectra and soliton families in multi-dimensions",
Opt. Lett. 39, 1133-1136 (2014).

\end{thebibliography}
\end{document}